\documentclass[aps,english,prl,floatfix,amsmath,superscriptaddress,tightenlines,twocolumn]{revtex4-2}
\usepackage{float}
\usepackage{graphicx}  
\usepackage{xcolor}
\usepackage{bm}        
\usepackage{braket}
\usepackage{amssymb}   
\usepackage{epstopdf}
\usepackage{algorithm} 
\usepackage{algpseudocode} 
\usepackage{xfrac}
\usepackage{cancel}
\usepackage{soul}
\usepackage{amsmath}
\usepackage{amsthm}
\usepackage{amsfonts}
\usepackage{bbm}
\usepackage{comment}
\usepackage{dsfont}
\usepackage{physics}
\usepackage{enumitem}

\usepackage[utf8]{inputenc}
\usepackage[english]{babel}
\usepackage[T1]{fontenc}
\definecolor{darkblue}{rgb}{0.1,0.2,0.6}
\definecolor{darkred}{rgb}{0.8,0.1,0.2}
\definecolor{darkgreen}{rgb}{0.31,0.62,0.24}
\definecolor{bleudefrance}{rgb}{0.19, 0.55, 0.91}
\usepackage[colorlinks,citecolor=darkblue,linkcolor=darkblue,urlcolor=darkblue]{hyperref} 
\usepackage{setspace}
\usepackage{url}  
\usepackage{mathrsfs}
\usepackage{mathtools} 
\usepackage{amsmath,amsthm,amsfonts,amssymb,xcolor}
\usepackage{orcidlink}

\begin{document}

\title{Probing mixed-state phases on a quantum computer via \\ Renyi correlators and variational decoding}

\author{Yuxuan Zhang\orcidlink{0000-0001-5477-8924}}
\email{quantum.zhang@utoronto.ca}
\affiliation{Department of Physics and Centre for Quantum Information and Quantum Control, University of Toronto,
60 Saint George St., Toronto, Ontario M5S 1A7, Canada}
\affiliation{Vector Institute, W1140-108 College Street, Schwartz Reisman Innovation Campus
Toronto, \\Ontario
M5G 0C6, Canada}
\author{Timothy H. Hsieh}
\email{thsieh@perimeterinstitute.ca}
\affiliation{Perimeter Institute for Theoretical Physics, Waterloo, Ontario N2L 2Y5, Canada}
\author{Yong Baek Kim}
\email{ybkim@physics.utoronto.ca}
\affiliation{Department of Physics and Centre for Quantum Information and Quantum Control, University of Toronto,
60 Saint George St., Toronto, Ontario M5S 1A7, Canada}
\author{Yijian Zou\orcidlink{0000-0001-8573-000X}}
\email{yzou@perimeterinstitute.ca}
\affiliation{Perimeter Institute for Theoretical Physics, Waterloo, Ontario N2L 2Y5, Canada}

\date{\today}

\begin{abstract}

Recent advances have defined nontrivial phases of matter in open quantum systems such as many-body quantum states subject to environmental noise. In this work, we experimentally probe and characterize mixed-state phases on Quantinuum's H1 quantum computer using two measures: Renyi correlators and the coding performance of a quantum error-correcting code associated with the phase. As a concrete example, we probe the low-energy states of the critical transverse field Ising model under different dephasing noise channels. First, we employ shadow tomography to observe a newly proposed Renyi correlator in two distinct phases: one exhibiting power-law decay and the other long-ranged.
Second, we investigate the decoding fidelity of the associated quantum error-correcting code using a variational quantum circuit, 
and we find that a shallow circuit is sufficient to distinguish the above-mentioned two mixed-state phases through the decoding performance quantified by entanglement fidelity. Our work is a proof of concept for the quantum simulation and characterization of mixed-state phases. 
 
\end{abstract}

\maketitle

\textit{Introduction.--}
Classifying quantum phases is a central endeavor in many-body physics which has largely focused on ground states. However, in real quantum devices, quantum states are far from isolated and pure, and recent works have explored nontrivial phases of matter in open quantum systems~\cite{verstraete2009quantum, diehl2008quantum, hastings2011nonzero, coser2019classification, fan2024diagnostics, bao2023mixed, Lee2023quantum, zou2023channeling, de2022symmetry, ma2023average, zhang2022strange, ma2023topological, rakovszky2024defining, sang2024mixed, markov, lu2023mixed, lee2022decoding, zhu2022nishimoris, chen2024symmetry, chen2024separability, lessa2024mixedstate, chen2024unconventional,lee2025symmetry, lee2024exact, sohal2024noisy, ellison2025toward, wang2025intrinsic, wang2024topologically, wang2025anomaly, xue2024tensor, guo2024locally, ma2025symmetry,su2024conformal,min2025mixed,lessa2025strong,kuno2024strong,ando2024gauge,kim2024persistent,molignini2023topological,weinstein2025efficient,su2024higher,min2024role,lee2024mixed,su2024emergent}. When a system undergoes a dissipative process, such as contact with a thermal bath or exposure to quantum noise, the resulting mixed state can exhibit phase transitions that extend beyond traditional pure-state and thermal paradigms.
As is the case with gapped and gapless ground state phases, for mixed states one can also have two varieties depending on whether correlations are short-ranged or critical.  

Critical mixed state phases, which can be obtained by decohering long-range entangled states \cite{Lee2023quantum,zou2023channeling,zhu} or measurement and feedforward \cite{lu2023mixed, zhang2025universalpropertiescriticalmixed}, 
constitute an important frontier in open quantum systems.  In particular, they are associated with novel types of quantum error correcting (QEC) codes \cite{sang2024approximate, yi2024complexity}. Traditional QEC codes, typically based on the stabilizer formalism and exemplified by topological codes, require implementation in at least two spatial dimensions. On the other hand, recent theoretical studies have shown that many physical states, even in one dimension, can be used as QEC codes. These physics-inspired QEC codes, such as ``CFT codes'' based on quantum critical states described by conformal field theory (CFT), exhibit coding properties that extend beyond those of conventional stabilizer codes~\cite{yi2024complexity,sang2024approximate}.  Mixed state phases and QEC properties are fundamentally connected ~\cite{sang2024mixed,sang2024approximate}: whether a noisy long-range entangled state remains in the same phase as the pure initial state is related to whether logical information encoded in the long-range entanglement can be decoded despite the noise.  
Nevertheless, it has been challenging to observe mixed-state phases and their code properties experimentally because they are typically characterized by nonlinear observables and information-theoretic quantities.  

In this work, we use two novel diagnostics to observe mixed-state phases associated with quantum criticality on a trapped-ion quantum processor~\cite{pino2021demonstration}. The first one is the \textit{Renyi correlators}. 
Specifically, we prepare quantum critical states described by a CFT subject to dephasing channels on the processor.  We then use randomized measurement techniques~\cite{huang2020predicting,elben2020mixed} to estimate Renyi correlators which distinguish mixed-state phases induced by 
different dephasing protocols, and we find results consistent with CFT calculations~\cite{Lee2023quantum}.

While Renyi correlators diagnose phase transitions associated with a given Rényi index (2 and above), they may not diagnose ``intrinsic'' phase transitions associated with von Neumann quantities. Thus, to complement Renyi correlators, we utilize a second diagnostic known as 
 \textit{entanglement fidelity}~\cite{schumacher1996quantum}, which is motivated by quantum error correction~\cite{peres1985reversible,shor1995scheme,gottesman1997stabilizer}. Essentially, entanglement fidelity quantifies the maximum recoverable information of an encoded quantum state after noise exposure.

\begin{figure*}
    \centering
    \includegraphics[width=.7\linewidth]{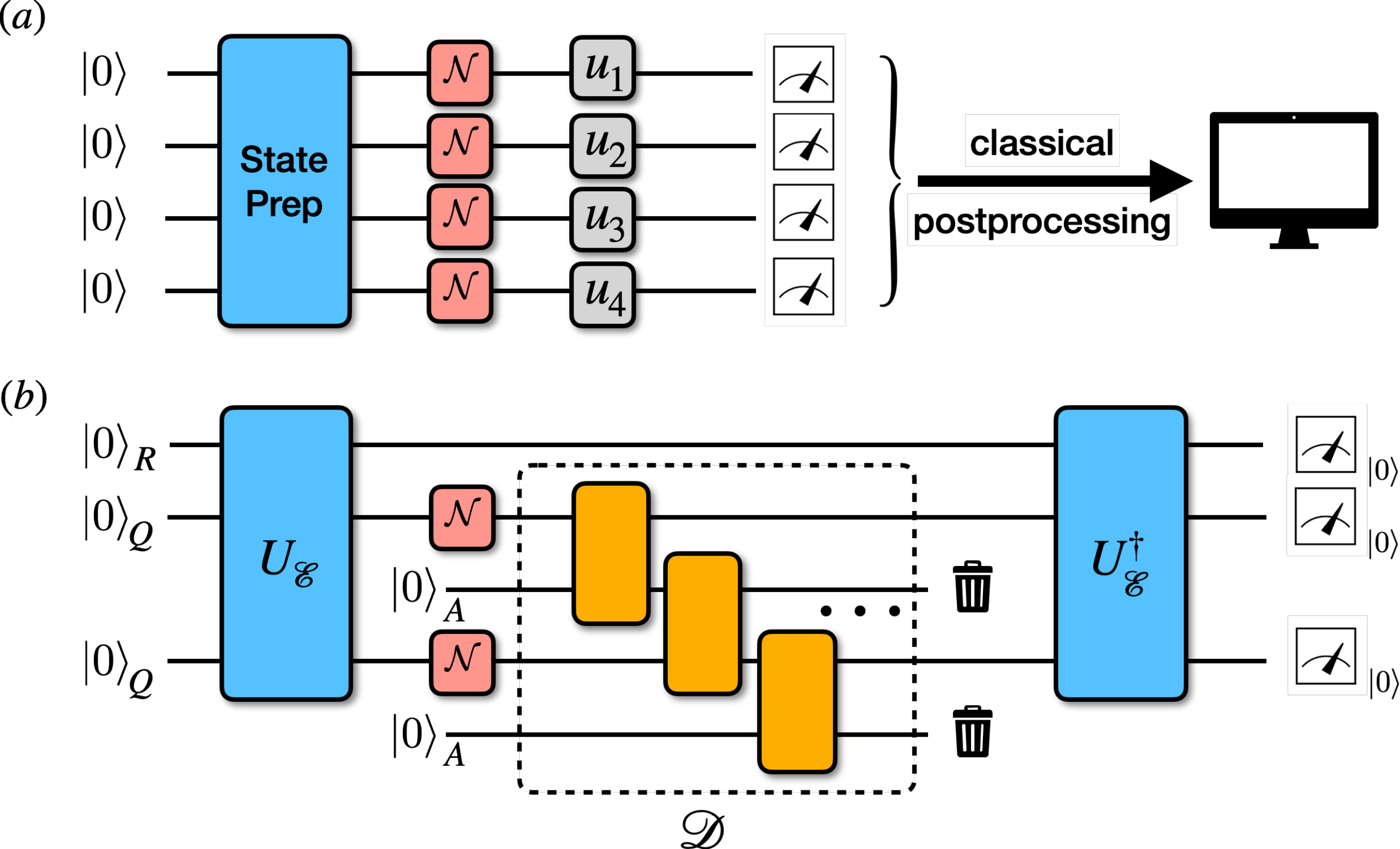}
    \caption{{\bf Circuit representation for two experimental setups to distinguish critical mixed state phases. }(a) We prepare a quantum critical state described by a CFT, apply the noise channel and compute the Renyi correlator with shadow tomography, see SM for more details. (b) We regard the CFT as a QECC with its ground state and first excited state as the codewords and diagnose mixed-state phases using the QEC property. The decoder is implemented as a variational circuit with ancilla, as shown in the diagram. The circuit is optimized to maximize the entanglement fidelity $F_e$.}
    \label{fig:fid_cir}
\end{figure*}

Depending on the nature of the dephasing channels, the threshold beyond which information is not recoverable may be zero or finite. In the critical transverse field Ising spin model in one dimension, a previous work \cite{sang2024approximate} showed that while $X$ dephasing has zero threshold, $Y$ and $Z$ dephasing have a finite decoding threshold, and the number of logical qubits may increase with the number of physical qubits. However, it has remained elusive how to find a decoding algorithm and how the code performs on an actual quantum computer.

We construct a decoder algorithm for CFT codes and perform a numerical study using variational optimization methods. By introducing ancilla qubits and formulating the decoder as a variational quantum circuit, our simulations and experimental observations reveal that entanglement fidelity indicates distinct decoder characteristics in the presence of $X$ and $Z$ errors, thereby confirming a theoretical prediction in \cite{sang2024approximate}. Consequently, 
both the Rényi correlator and entanglement fidelity confirm the experimental observation of two distinct mixed-state phases at quantum criticality at both Renyi index 2 and the von Neumann limit \footnote{We note that Z dephasing corresponds to different kinds of perturbations in the renormalization group (RG) flow depending on the Renyi index: this perturbation is marginal for Renyi index 2 and irrelevant for the von Neumann limit \cite{sang2024approximate}}.

\textit{Setup and Renyi correlators. --} 
We consider the ground state $|\psi\rangle$ of the critical transverse-field Ising model $$H = - \sum_{i=1}^L X_i X_{i+1} - \sum_{i=1}^L Z_i$$  
 subject to $\mathcal{N}^{[i]}$~\cite{zou2023channeling}, the dephasing channels in $X$ or $Z$ directions. In Ref.~\cite{zou2023channeling}, it was shown that the different negativity scaling between two mixed states indicates a mixed-state phase transition. 
In this work, we show that the distinction also manifests in the Rényi correlators for $n\geq2$ (which permits more efficient experimental detection).  Given a mixed state $\rho$, we define the $n-$th Renyi observables through
\begin{equation}
    \langle O \rangle^{(n)} = \frac{\tr (\rho^n O)}{\tr(\rho^n)},
\end{equation}
where $O$ is an operator and $n$ is a positive integer (not to be confused with a similar quantity studied in~\cite{Lee2023quantum,sala2024spontaneous,lessa2025strong} to detect spontaneous strong to weak symmetry breaking). The Renyi observable is the expectation value of $O$ on the normalized state $\rho^n/\tr(\rho^n)$ 
and reduces to the usual expectation value $\langle O\rangle^{(1)}=\tr(\rho O)$ for $n=1$. We will consider the Renyi two-point correlator, $C^{(n)}_{O_1 O_2}(l):= \langle O^{[0]}_1O^{[l]}_2\rangle^{(n)}$, where $O_1$ and $O_2$ are local operators on site $0$ and site $l$, respectively. 

\begin{figure}
    \centering
    \includegraphics[width=1\linewidth]{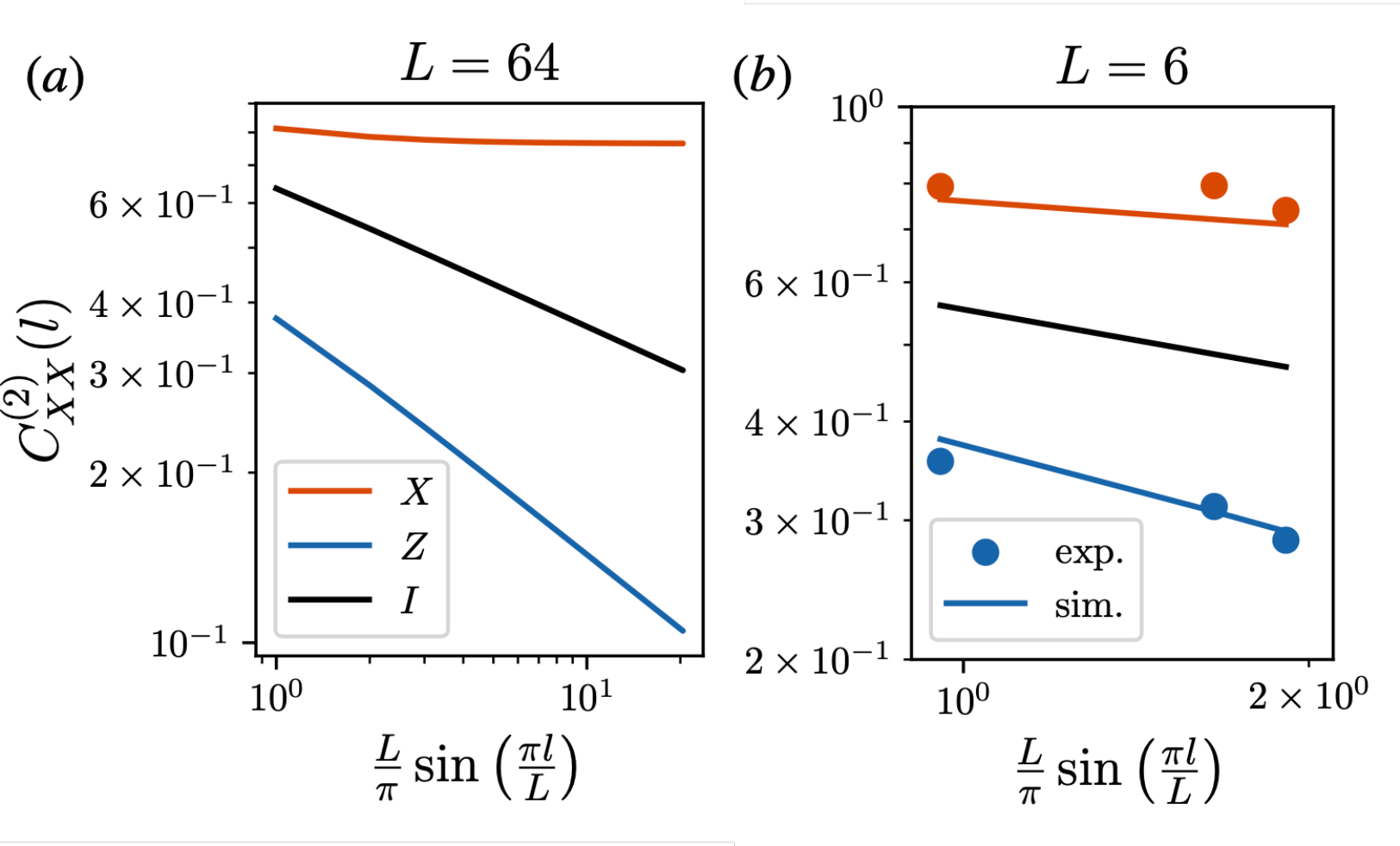}
    \caption{{\bf Renyi correlators.} (a) $C^{(2)}_{XX}$ for the Ising ground state under different dephasings, with total system size $L=64$ and $p=0.3$. The exponents for $Z$ and $I$ dephasings are $\eta^{Z}_2 = 0.35(1)$ and $\eta^{I}_2 = 1/4$, while for $X$ dephasing the correlator saturates to a constant. (b) Experimental results on the Quantinuum H1 trapped-ion processor. A jackknife estimation reports that the error bars are less than 1 percent, showing that the deviation is mostly due to device noise. The experimental estimation for the slope is $\approx 0.31$ for $Z$ and $\approx 0.07$ for $X$ due to finite-size corrections and noise.}
    \label{fig:XX}
\end{figure}

In the thermodynamic limit, the Renyi $XX$ correlator at long distances takes the form of,
\begin{equation}
    C^{(n)}_{XX}(l) = A^{(n)} + B^{(n)} \left( \frac{L}{\pi}\sin\left(\frac{\pi l}{L}\right)\right)^{-\eta_n},
\end{equation}
where $A^{(1)} = 0$ and $\eta_1 = 1/4$ is the critical exponent of the Ising model for both dephasing channels. The correlator shows distinctive properties for the $X$ and $Z$ channels at $n\geq 2$. For the $X$-dephased state, the correlator becomes long-range, i.e., $A^{(n)}> 0$. For the $Z$-dephased state $A^{(n)}=0$ and $\eta_n$ decreases continuously for $n\geq 2$. This is verified through the numerical simulation on $L=64$ spins shown in Fig.~\ref{fig:XX}(a).  

The result can be explained through the mapping from decoherence channels to conformal defects \cite{zou2023channeling,Lee2023quantum,ma2023exploringcriticalsystemsmeasurements}. The exponents $\eta_2$ give exactly twice the scaling dimensions of defect operators in the replicated CFT. For $X$ dephasing, it induces a relevant spin-spin coupling on the defect, making the defect long-range ordered such that $A^{(2)}>0$. The $Z$ dephasing is a marginal perturbation, such that $\eta_2$ continuously changes with dephasing rate $p$. 
Interestingly, our numerical fit suggests that $\eta_2 \approx 5/4$ for the subleading exponent in $C^{(2)}_{XX}$ under $X$ dephasing, whereas the naive expectation that the defect is factorized into a direct sum of polarized boundaries predicts that $\eta_2 = 4$. We also computed other Renyi correlators, which show interesting new exponents, see SM for details.

\textit{Shadow tomography. --} 
We use the randomized measurement method~\cite{elben2020mixed} known as shadow tomography to reconstruct Renyi correlators from experimental data. 
One key advantage of shadow tomography is that, after measuring and recording the data, one can extract various non-commuting quantum information quantities, such as the Rényi correlator, from the same data set. We briefly review the method here: For each copy of the state, we first perform a randomized measurement by applying single-qubit unitaries $u_{1} \otimes u_2\otimes\dots \otimes u_{L} $, each sampled independently from an ensemble that forms a unitary 3-design~\cite{dankert2009exact,elben2020mixed} 
and record the classical measurement outcomes $\bm{k} = \{k_1, \dots, k_L\}$. The experiment is repeated $ M $ times, and we denote each set of measurement outcomes as $ \bm{k}^{(r)} $, from which we can construct an unbiased estimator of the density matrix. 

\begin{align}
    \hat{\rho}^{(r)} = \bigotimes_{i=1}^{L} \hat{\rho}^{(r)}_{i} 
    = \bigotimes_{i=1}^{L} (3 u_i^{\dagger}\ketbra{\bm{k}^{(r)}_i}{\bm{k}^{(r)}_i}u_i - \mathcal{I}),
\end{align}
whose expectation gives raise to $\rho$~\cite{huang2020predicting}. 
For $n = 2$ Renyi observables, we find that

\begin{align}
    \langle \tr (\rho^2 O)\rangle = \frac{1}{2!}{M\choose 2}^{-1} \tr{\sum_r[\hat{\rho}^{(r)}]^2O - \sum_r[(\hat{\rho}^{(r)})^2])O},
\end{align}

To increase the quality of data further, we use the system's translational symmetry. Namely, we permute the qubits and construct independent estimations of $C^{(2)}_{XX}(l)$ and report their mean value as the estimation of $C^{(2)}_{XX}(l)$.

The experimental procedure is as follows:
\begin{enumerate}
    \item Prepare the critical Ising ground state with a variational quantum circuit.
    \item Apply the dephasing quantum channel: a Pauli operator is applied to each site with probability $p$.
    \item Apply a random single qubit rotation $u_i$, independently sampled from a 3-design, to each physical qubit.
    \item Measure in the computational basis and record the measurement result.
    \item Repeat 1-4 until the desired data set size is reached.
    \item Classical post-process to get an estimation of the quantum information quantities.
\end{enumerate}
We implement this experimental procedure for both the X and Z dephasing with a strength of \( p=0.3 \)  on Quantinuum's H1 trapped-ion quantum computer (with 40,000 circuit runs each) and report the measured Renyi correlators in Fig.~\ref{fig:XX}~\footnote{In SM, we also compute the Renyi negativity and find agreement with the theoretical expectations.}. Our results indicate that the experimental setup recovers Rényi correlators within an inaccuracy of $\sim10\%$ compared to exact diagonalization (ED) (see also SM for a shadow estimation of Renyi negativity).

We observe clear distinctions between the state under different channels, i.e., X dephasing, Z dephasing, and identity channel. In terms of the amplitude of $C^{(2)}_{XX}(l)$, we observe a significant increase under $X$ dephasing and a significant decrease under the $Z$ dephasing. Furthermore, we can approximately extract the slope of the decay of $C^{(2)}_{XX}(l)$ on a logarithmic scale. We find the slopes at $\approx 0.31$ for $Z$ and $\approx 0.07$ for $X$ dephasing, respectively. Compared with the identity channel with slope at $0.25$, we observe the $Z$ dephasing increases the exponent and the $X$ dephasing decreases the exponent, in accordance with the theoretical prediction. These findings suggest that even at this relatively small system size, signatures of two distinctive mixed-state phases induced by noise are observed on a quantum computer.

\textit{Decoding a CFT code--} Quantum criticality can also serve as a QEC code, whose decodability can be used as 
another way to observe the aforementioned two mixed-state phases in the low-energy subspace for the critical spin chain with Hamiltonian $H$. Specifically, one 
could encode $D$ logical states using the low-energy subspace spanned by the $D$ lowest-energy eigenstates, $|\phi_{\alpha}\rangle$ of $H$. This so-called CFT code may have a finite decoding threshold $p_c \neq 0$ in the thermodynamic limit, albeit at finite sizes the code becomes approximate~\cite{yi2024complexity,sang2024approximate}. In particular, the Ising CFT code can correct uniform $Z$ dephasing error for $p<1$, but has a zero decoding threshold for the $X$ dephasing error. 

Formally, let $R$ be a $D$-dimensional reference qudit and the code subspace spanned by $\{|\phi_{\alpha}\rangle_Q, ~ 0\leq \alpha \leq D-1\}$ and the maximal entangled state $|\phi_{RQ}\rangle= \frac{1}{\sqrt{D}}\sum_{\alpha=0}^{D-1} |\alpha\rangle_R |\phi_{\alpha}\rangle_Q$, where $Q$ is the set of physical qubits. Throughout the rest of the paper, we set $D=2$ for consistency. Let $\mathcal{N}_Q = \otimes_{j\in Q} \mathcal{N}^{[j]}$ be the noise channel acting on $Q$ and we can define the noisy state
\begin{equation}
    \rho_{RQ} = \mathcal{N}_Q(|\phi_{RQ}\rangle\langle \phi_{RQ}|)
\end{equation}
We will use \textit{entanglement fidelity}, $F_e$, to quantify the decodability of the CFT code, which is defined as: 
\begin{equation}
    F_e = \max_{\mathcal{D}} \langle \phi_{RQ}|\mathcal{D}_Q(\rho_{RQ}) |\phi_{RQ}\rangle.
    \label{eq:fe}
\end{equation}
where $\mathcal{D}_Q$ is a decoding channel acting on $Q$. 
Since the explicit construction of decoders is unknown, we propose a greedy algorithm to compute $F_e$ for the CFT code at small system sizes and compare it with the bounds defined below.
Specifically, we introduce ancilla qubits, $A$, and dilate the channel $\mathcal{D}_Q$ into an isometry $W: Q\rightarrow QA$ such that $\mathcal{D}_Q(\rho) = \tr_A W \rho W^{\dagger}$. The entanglement fidelity can then be written as 
\begin{equation}
    F_e(W,W^{\dagger}) = \langle \phi_{RQ}|\tr_A\{W \rho_{RQ}W^{\dagger}\}|\phi_{RQ}\rangle.
\end{equation}

The optimization is as follows: at each step, we keep $W^{\dagger}$ fixed and compute the maximum of $F_e(W, W^{\dagger})$ over $W$, and then compute the maximum over $W^{\dagger}$ by fixing $W$. The step is repeated until $F_e$ converges. We typically find that $F_e$ converges in less than $5$ iterations, see SM for details.

Optimizing $F_e$ for large systems soon becomes formidable, but one may bound $F_e$ through various quantities that are easier to compute. One quantity is the channel distance, defined as
\begin{equation}
\label{eq:drho}
    d_\rho = \sqrt{1 - f(\rho_{RE}, \rho_R\otimes \rho_E)},
\end{equation}
where $\rho_{RE} = \tr_{Q} \rho_{RQE}$. Here $\rho_{RQE}$ is a purification of the state $\rho_{RQ}$ with environment $E$ coming from the dilation of the noise channel $\cal N$, and $f$ represents the mixed-state fidelity function.  
Intuitively, this means decoding is possible if and only if the noise process leaves the $E$ uncorrelated with the $R$, i.e., no logical information is lost to the environment~\cite{knill1997theory,devetak2005private}.
The channel distance is independent of the choice of purification and provides a lower and upper bound on $F_e$ in the following way \cite{Beny_2010}:
\begin{equation}
\label{eq:boundFe}
   \frac{1}{2} d_{\rho}\leq \sqrt{1-\sqrt{F_e}}\leq d_\rho
\end{equation}

\begin{figure}
    \centering
    \includegraphics[width=0.8\linewidth]{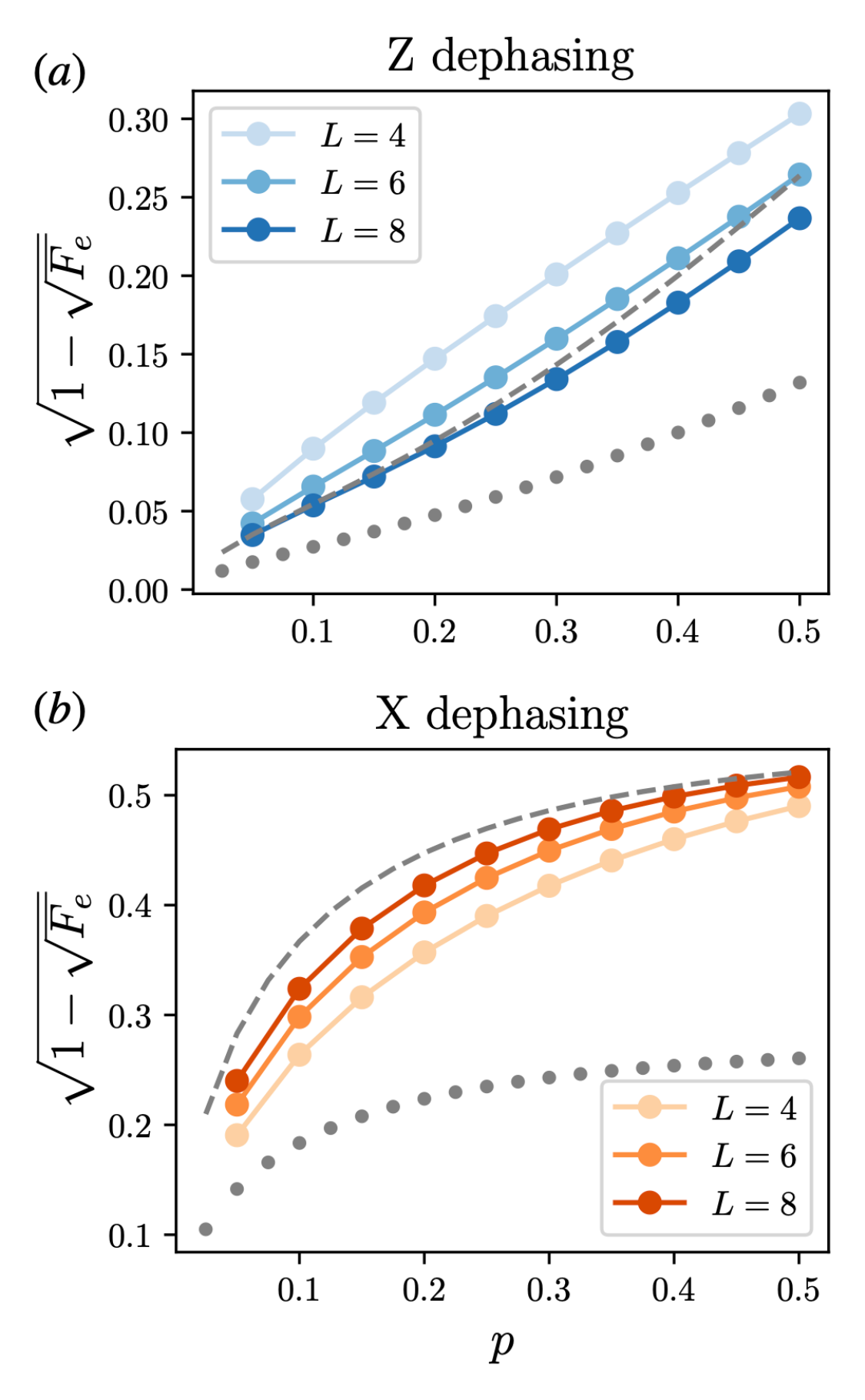}

    \caption{\textbf{Optimal entanglement fidelty.} Optimal entanglement fidelity $F_e$ of the Ising CFT code under $X$ and $Z$ dephasing versus channel distance $d_{\rho}$ in Eq.~\eqref{eq:drho}. The dotted and dashed grey lines represent $d_\rho$ and $d_\rho/2$ at $L=8$, which are upper and lower bound for $\sqrt{1-\sqrt{F_e}}$, respectively. The upper bound $d_{\rho}$ is saturated for $Z$ and $X$ dephasings at small and large $p$, respectively.}
    \label{fig:Fe_optimial}
\end{figure}

For $L=4,6,8$, we compare the numerically optimized $F_e$ to this bound in Fig.~\ref{fig:Fe_optimial}. The optimized entanglement fidelity $F_e$ is consistent with the bound Eq.~\eqref{eq:boundFe} and saturates the upper bound for weak $Z$ dephasing. Additionally, for the $Z$ dephasing, $F_e$ increases with $L$ for fixed $p$, indicating a finite decoding threshold, as logical information is better preserved by adding more physical qubits. For the $X$ dephasing, $F_e$ decreases with $L$ for fixed $p$, indicating zero decoding threshold. Nevertheless, we note that the CFT code can still correct subextensive number of $X$ errors if $p<O(L^{-3/4})$. As we will see, this large gap between the number of correctable errors in $X$ and $Z$ dephasing is useful for the experimental detection of the two mixed-state phases using variational decoding.

\begin{figure}
    \centering
    \includegraphics[width=1.\linewidth]{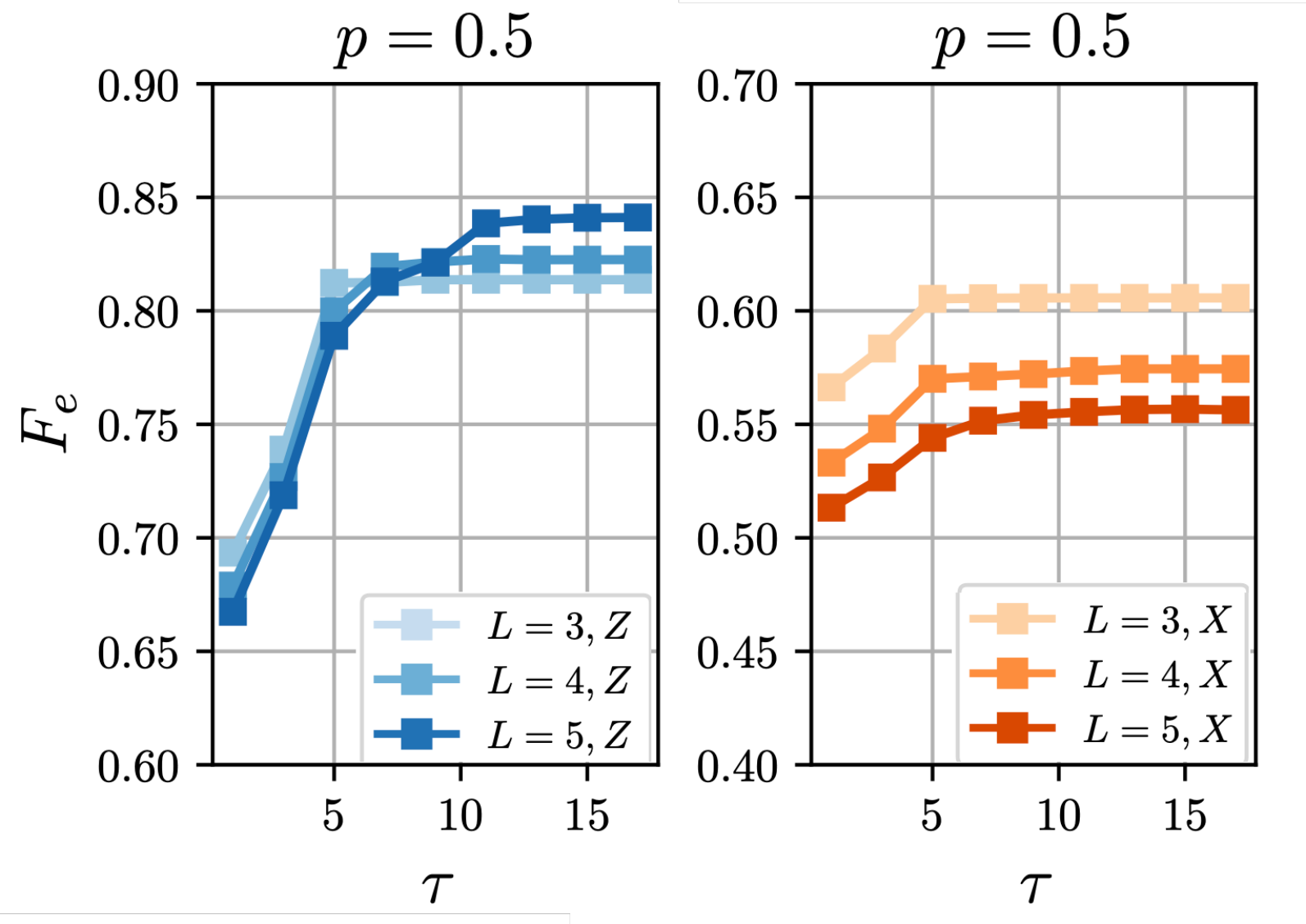}
    \caption{\textbf{Entanglement fidelity from variational decoding}. Fixing $L=3, 4, 5$ and constant noise strength $p=0.5$, we numerically optimize $F_e$ at various decoder circuit layers up to $\tau = 17$ with a gradient decent optimizer. We observe that for larger $L$, the optimal decoder for $Z$ dephasing requires more layers and achieves better $F_e$, while the optimal $X$ decoder requires much fewer layers, and the saturation $F_e$ decreases with $L$.}  
    \label{fig:fid_sim}
\end{figure}
\textit{Variational decoding. --} In order to implement the decoder on a quantum computer, we resort to a variational approach where the channel is dilated into a variational unitary circuit with ancillas \footnote{According to the HJW theorem, any mixed state can be purified with twice the number of qubits. Thus, the mapping between two such states can be implemented with a unitary operation with one ancilla qubit per system qubit.}: We introduce an ancilla qubit between each nearest-neighbor pair of system qubits and construct a ladder-shaped variational quantum circuit. This circuit structure is first proposed as a quantum realization of the matrix product state~\cite{schon2005sequential,perez2006matrix,schon2007sequential,foss2021holographic,niu2022holographic,zhang2022qubit,anand2023holographic,zhang2024sequential}, and is illustrated in Fig.~\ref{fig:fid_cir}b. The ladder structure is more expressive than brick-wall structure when comparing circuits with the same gate count, as noted in recent works~\cite{chen2024sequential,zhang2025observation}. For Quantinuum H1 processor's architecture, the main error source comes from entangling gates, this reduced gate count makes the sequential circuit more favorable (see SM for more details).

In order to benchmark the performance of variational decoding, we choose three small physical system sizes, $L=3,4,5$ with noise strength $p=0.5$ and optimize the variational circuit up to $\tau=17$ layers. 
The result is shown in Fig.~\ref{fig:fid_sim}, where two observations can be made. (1) To begin with, comparing the $L=5$ to $L=3, 4$, the variational decoder saturates to a larger $F_e$ in the presence of $Z$ dephasing, whereas the saturation $F_e$ decreases for $X$ dephasing, matching our theoretical prediction. (2) The decoder performance for $X$-noise nearly saturates at $\tau=5$ for all three sizes, whereas for the $Z$-dephasing, the saturation depths grow substantially with $L$, which are 5, 7, 11 for $L=3,4,5$, respectively. 
\begin{figure}
    \centering    \includegraphics[width=.9\linewidth]{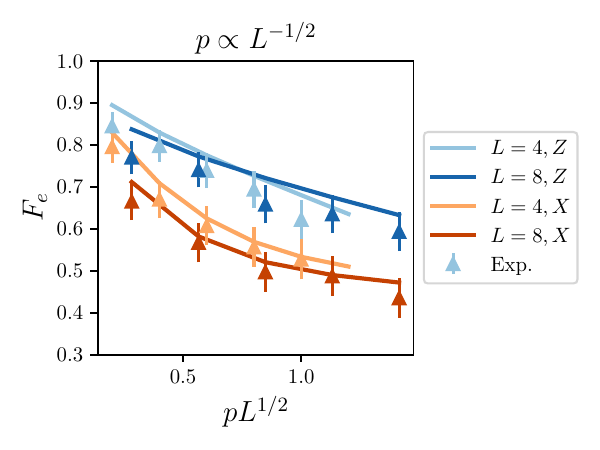}
    \caption{\textbf{Experimental results on variational decoding}. Experimental results on entanglement fidelity $F_e$ of the Ising CFT code with $L=4$ and $L=8$ with a single-layer decoder (1000 shots), with $p\propto L^{-1/2}$. Error bars represent three times the standard deviation due to finite sampling. For $Z$ dephasing at large $p$, $F_e$ increases with $L$, which persists even with device noise. Whereas for $X$ dephasing, $F_e$ strictly decreases with $L$ even if the error rate $p\propto L^{-1/2}$.}     \label{fig:fid_exp}
\end{figure}

In the experiment, we fix $L=4,8$ and $d = 1$ and set the error rate to be $p\propto L^{-1/2}$. 
This noise rate is correctable for $Z$ dephasing but not for $X$ dephasing. Even at decoding circuit depth 1, the output fidelity already reveals the difference between the mixed-state phases induced by the two quantum channels: In Fig.~\ref{fig:fid_exp}, for $Z$ dephasing, there exists a region ($pL^{1/2}\gtrapprox1)$ where the $L=8$ code outperforms the $L=4$, whereas for the $X$ dephasing the $L=4$ code is strictly better than $L=8$. This distinction is captured by the experimental data even in the presence of additional physical noise 
(for $L=8$ we estimate the physical noise introduces a $\sim9\%$ deviation; see SM for more details). Interestingly, the experimentally measured $F_e$ for the $Z$ dephasing is less accurate than the $X$ dephasing when compared with the simulation because of the presence of the physical $X$-error.

Compared with Renyi observales, which requires $\exp(O(L))$ samples for shadow tomography~\footnote{Measuring Renyi correlator with SWAP test also needs exponential sampling as the purity is exponentially small in $L$}, the sampling complexity for variational decoding is $O(1)$. Unless the decoder requires an exponential number of gates, the variational decoding is much more efficient than measuring Renyi observables on a quantum computer. 
\textit{Discussion.--}  
In this work, we have demonstrated that mixed-state phases in critical quantum systems can be observed experimentally using two complementary diagnostic tools: Renyi correlators and variational decoding. We employed shadow tomography to measure second-order Rényi two-point correlators that revealed distinct power-law behaviors under different dephasing channels. The other approach, based on variational decoding, optimizes the entanglement fidelity of the QECC based on the low-energy subspace of the same quantum system, successfully distinguishing between the two error-correctiing phases induced by $X$ and $Z$ dephasing channels even at shallow circuit depths. 

While Rényi correlators require no ancilla qubits to measure, they detect phase transitions specific to integer Rényi index. On the other hand, entanglement fidelity probes the information-theoretic phase transition~\cite{sang2024approximate}, capturing transitions that are otherwise challenging to observe. This distinction makes these two measures complementary experimental tools for mapping out mixed-state phases in open quantum systems. Another advantage of the variational decoding approach as a diagnostic for mixed-state phases is that it overcomes the exponential sample complexity typically associated with Rényi correlators. The trade-off is the number of layers in the decoding circuit. Based on our numerical simulations on small sizes, we conjecture that optimally decoding a correctable dephasing error at a finite rate requires the circuit layer scaling as $\tau=O(L)$, whereas for an uncorrectable dephasing, the optimal decoding circuit depth grows much more slowly. This would suggest that detecting the correctable and uncorrectable phases can be achieved at polynomial cost on a quantum computer.

Looking ahead, our work opens several promising directions for future research. First, from a field-theoretic standpoint, it is intriguing to explore the universal properties of the decoherence-induced defect and to develop an analytical understanding of the associated Rényi correlators. A second open problem is the analytical study of optimal decoders for CFT-based quantum codes considered in this work. Third, techniques such as shadow tomography and variational decoding could be leveraged to experimentally observe other mixed-state phases, including symmetry-protected topological (SPT) phases and topologically ordered states. Lastly, by combining variational decoding with machine learning, it may be possible to discover error-correcting protocols capable of mitigating physical noise—even in cases where the noise model is unknown or ill-defined.
\paragraph{Acknowledgement}
We would like to thank Cenke Xu and Jeongwan Haah for their insightful discussions. Y.Zhang and Y.B.K. were supported by the Natural Science and Engineering Research Council (NSERC) of Canada. Y.Zhang and Y.B.K. acknowledge support from the Center for Quantum Materials at the University of Toronto. Y.Zhang was further supported by a CQIQC fellowship at the University of Toronto. This research used resources of the Oak Ridge Leadership Computing Facility, which is a DOE Office of Science User Facility supported under Contract DE-AC05-00OR22725, and is supported, in part, by grant NSF PHY-2309135 to the Kavli Institute for Theoretical Physics (KITP). This work was supported by the Perimeter Institute for Theoretical Physics (PI) and the Natural Sciences and Engineering Research Council of Canada (NSERC). Research at PI is supported in part by the Government of Canada through the Department of Innovation, Science and Economic Development Canada and by the Province of Ontario through the Ministry of Colleges and Universities.

\bibliographystyle{apsrev4-2}
\bibliography{ref}

\appendix
\newpage
\onecolumngrid
\onecolumngrid
\section{Different exponents in Renyi correlators}

In this section we present numerical simulations of Renyi correlators of the critical TFIM under $X$ or $Z$ noise with $L=64$ spins and fit for their critical exponents.

\subsubsection{Field-theoretic intepretation}
The mixed state $\rho$ is obtained through a pure state under local decoherence $\rho = \mathcal{N}(|\psi\rangle\langle\psi|)$, where $\mathcal{N}$ is a product of local channels, $\mathcal{N} = \otimes_j \mathcal{N}^{[j]}$. As one varies the parameters of the local channels, one observes distinct behaviors between Renyi correlators with $n=1$ and $n\geq 2$. The $n=1$ correlator can be related to two-point function of the original pure state through $C^{(1)}_{O_1 O_2}(l)=\langle O^{[0]}_1 O^{[j]}_2\rangle^{(1)}= \tr(\rho O^{[0]}_1 O^{[j]}_2) = \tr(\mathcal{N}(|\psi\rangle\langle \psi|)O^{[0]}_1 O^{[j]}_2 ) = \tr(\mathcal{N}^{*}(O^{[0]}_1 O^{[j]}_2)|\psi\rangle\langle \psi|) = \langle \psi  | \mathcal{N}^{[0]*}(O^{[0]}_1)\mathcal{N}^{[j]*}(O^{[j]}_2) |\psi\rangle$, whereas the $n\geq 2$ observables $\tr(\rho^n O^{[0]}_1 O^{[j]}_2) = \tr(\mathcal{N}^{\otimes n}((|\psi\rangle\langle\psi|)^{\otimes n}) \tau_n O^{[0]}_1 O^{[j]}_2) = \langle \psi^{\otimes n}|\mathcal{N}^{*\otimes n}(\tau_n O^{[0]}_1 O^{[j]}_2)|\psi^{\otimes n}\rangle$, which involves a string operator as $\mathcal{N}^{*\otimes n}(\tau_n)$ is a product of the twist operators on all sites, $\tau_n = \otimes_j \tau^{[j]}_n$. The string operator then corresponds to a defect in the $n-$copied CFT. 

\begin{figure}[t]
    \centering
    \includegraphics[width=0.85\linewidth]{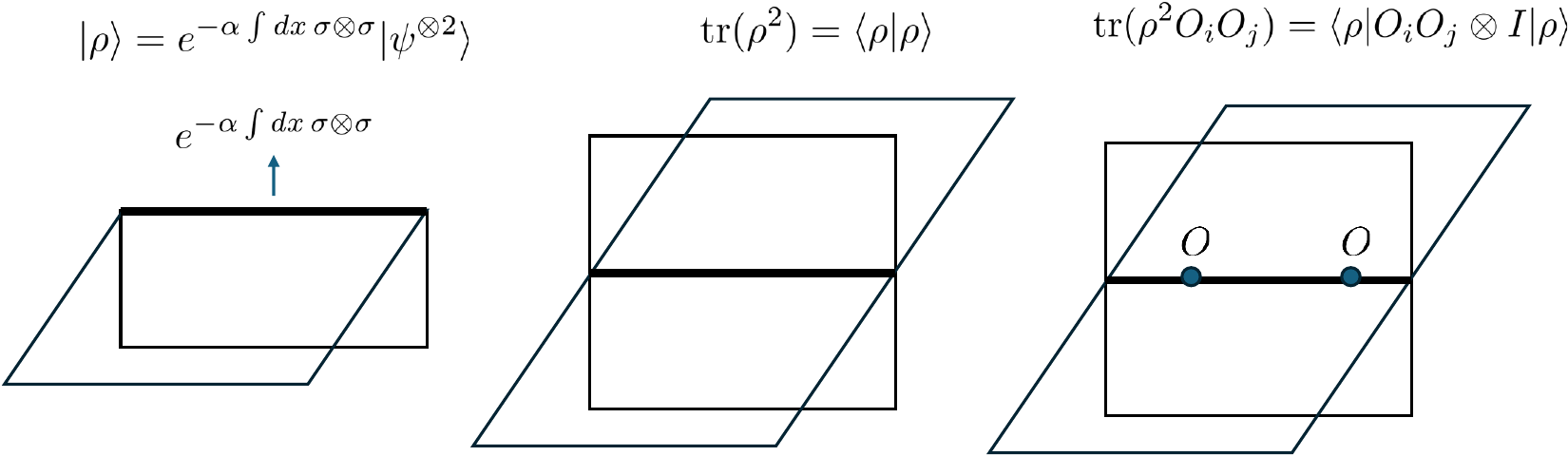}
    \caption{Taking the critical transverse-field Ising model under $X$ dephasing as an example, we illustrate the path integral representations of the doubled state $|\rho\rangle$ (left), the purity $\tr(\rho^2)$ (middle) and the Renyi correlator $\tr(\rho^2 O_i O_j)$ (right). Each rectangle represents the path integral on half plane, preparing one copy of the ground state $|\psi\rangle$. The thick line, inserted at the $\tau=0$ slice, represents the defect that corresponds to the dephasing channel in the replicated CFT $\mathrm{Ising}^{\otimes 2}$.  The Renyi correlator corresponds to the two-point correlation function of defect operators.}
    \label{fig:path}
\end{figure}

For $n=2$, we can further regard the density matrix as a state on the doubled Hilbert space $|\rho\rangle$. Then the purity and the Renyi correlator are the usual norm and correlation function of the doubled state, i.e., $\tr\rho^2 = \langle \rho|\rho\rangle$ and $\tr(\rho^2 O_1 O_2) = \langle \rho| O_1 O_2 \otimes I |\rho\rangle$. The evolution of the doubled state under local channels can be represented as a short-time imaginary time evolution, which renormalizes into a line defect inserted at imaginary time $\tau = 0$ slice, see Fig.~\ref{fig:path}. The Renyi correlator becomes a two-point correlation function on the defect, as shown in Fig.~\ref{fig:path}. Under the conformal transformation onto a cylinder with circumference $L$ , one then expects that
\begin{equation}
    C^{(2)}_{O_1 O_2}(l) = A + B \left( \frac{L}{\pi}\sin\left(\frac{\pi l}{L}\right)\right)^{-\eta_2}
\end{equation}
where $\eta_2$ is twice the leading scaling dimension of the defect operator and $A$ represents a possibly nonzero constant piece.

Below we study the Renyi correlators for the critical Ising model under $X$ and $Z$ dephasing numerically and extract the critical exponents $\eta_2$. We will find a number of new exponents that do not appear in the defect operator content of a single-copied Ising CFT, where the latter is thoroughly studied in Ref.~ \cite{Oshikawa_1997}. This suggests that the decoherence corresponds to more intricate non-factorized defect of the double-copied Ising CFT, which is currently not fully classified.

\subsubsection{X dephasing}
\begin{figure}
    \centering
    \includegraphics[width=0.45\linewidth]{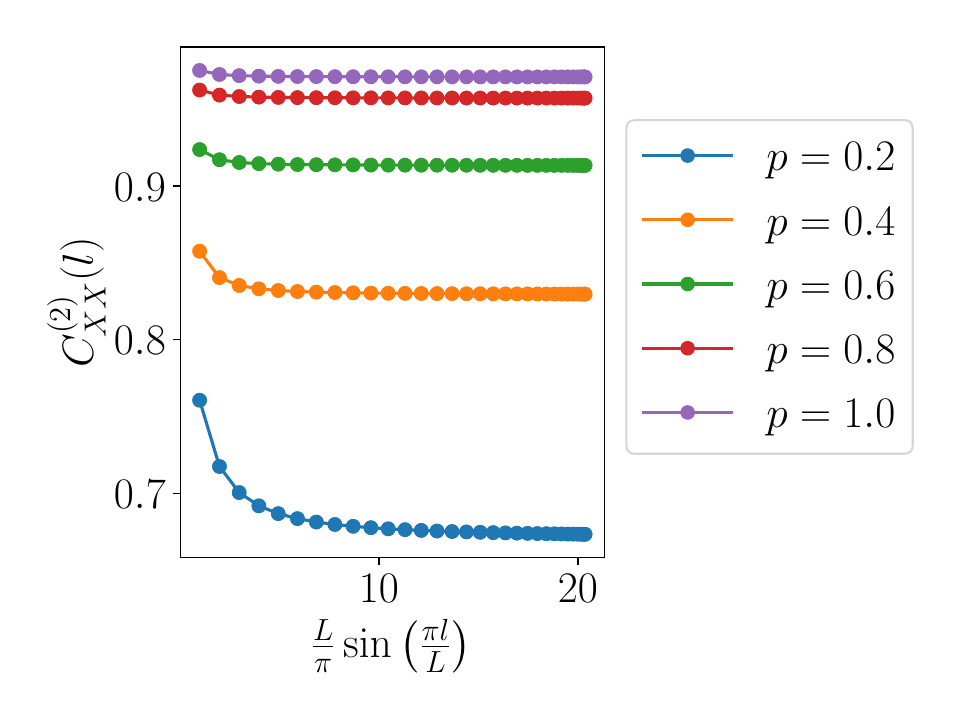}
    \includegraphics[width=0.45\linewidth]{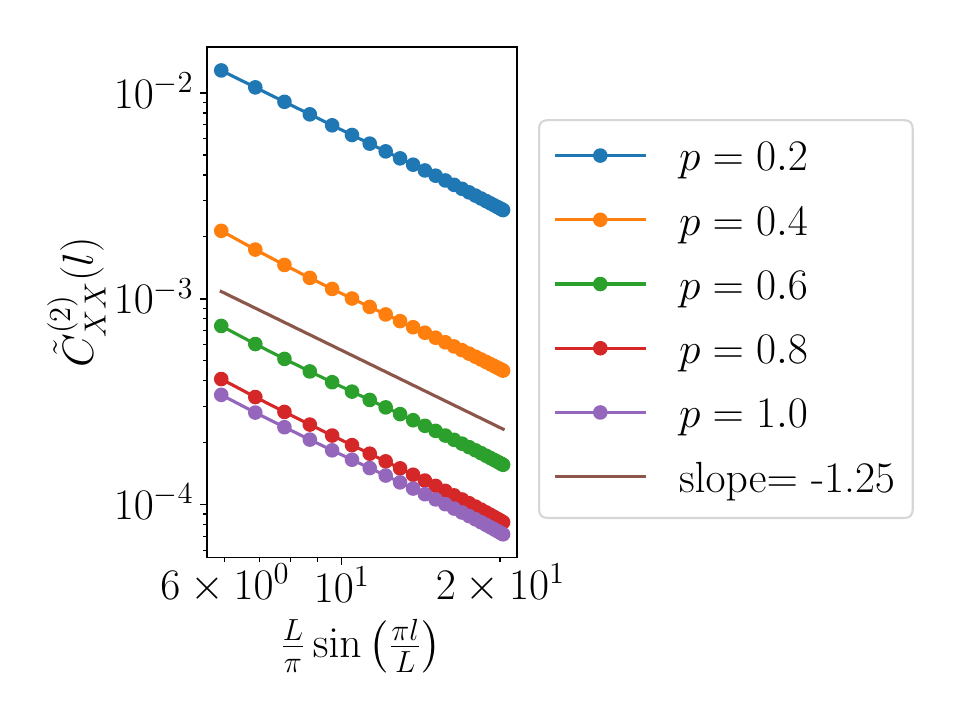}
    \caption{Renyi-2 XX correlator $C^{(2)}_{XX}(l)$ of the transverse field Ising ground state under X dephasing. For any $0<p\leq 1$ the correlator saturates to a constant at large $l$ (left). After subtracting the constant part, the tail is a power law decay with an exponent $\eta_2 \approx 1.25(5)$ (right).} 
    \label{fig:XX-X}
\end{figure}

We consider the Renyi correlators for $X$ dephasing.
\begin{equation}
    C^{(2)}_{XX}(l) = \tr(\rho^2 X_0 X_l)/\tr(\rho^2)
\end{equation}
The correlation function is long-ranged, see Fig.~\ref{fig:XX-X}. This suggests that the function is of the form 
\begin{equation}
    C^{(2)}_{XX}(l) = A + B \left( \frac{L}{\pi}\sin\left(\frac{\pi l}{L}\right)\right)^{-\eta_2},
\end{equation}
where $\eta_2$ is twice the scaling dimension of a defect operator. 
In order to compute $\eta_2$, we denote
\begin{equation}
    \tilde{C}^{(2)}_{XX}(l) = C^{(2)}_{XX} - A.
\end{equation}
We use the linear fit in logarithmic scale
\begin{equation}
    \log(C^{(2)}_{XX}(l)-A) = -\eta_2 \log \left( \frac{L}{\pi}\sin\left(\frac{\pi l}{L}\right)\right) + \log B
\end{equation}
such that the slope corresponds to $-\eta_2$. Note that $A$ is not universal.
We determine $A$ numerically such that the linear fit agrees with the numerical data best. The result is shown in Fig.~\ref{fig:XX-X}. We find that $\eta_2 \approx 1.25(5)$ independent of $p$. 

\begin{figure}
    \centering
    \includegraphics[width=0.45\linewidth]{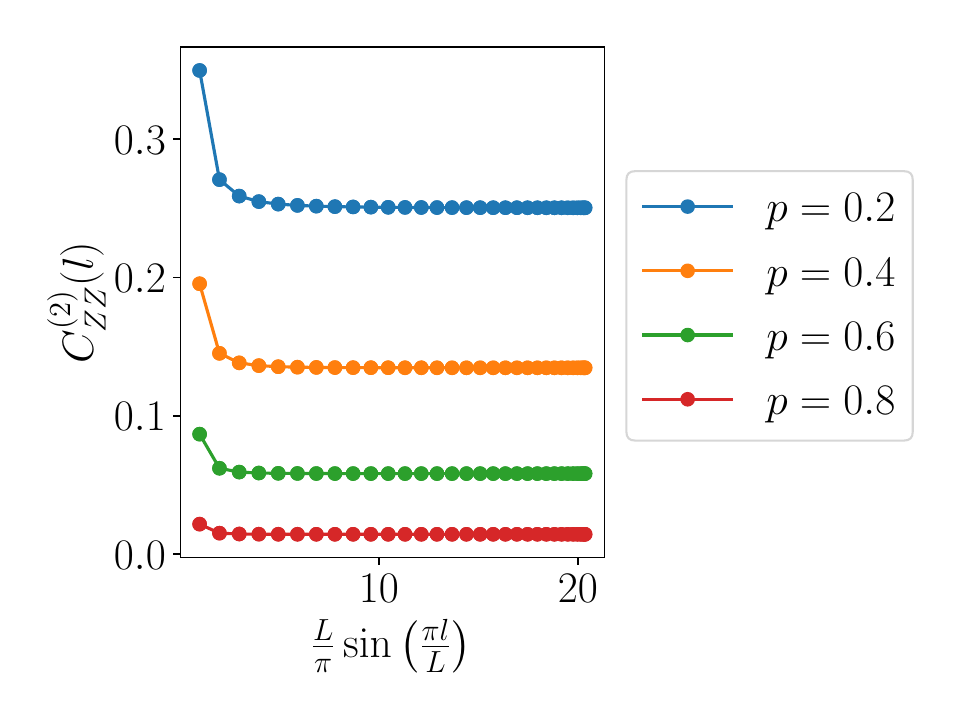}
    \includegraphics[width=0.45\linewidth]{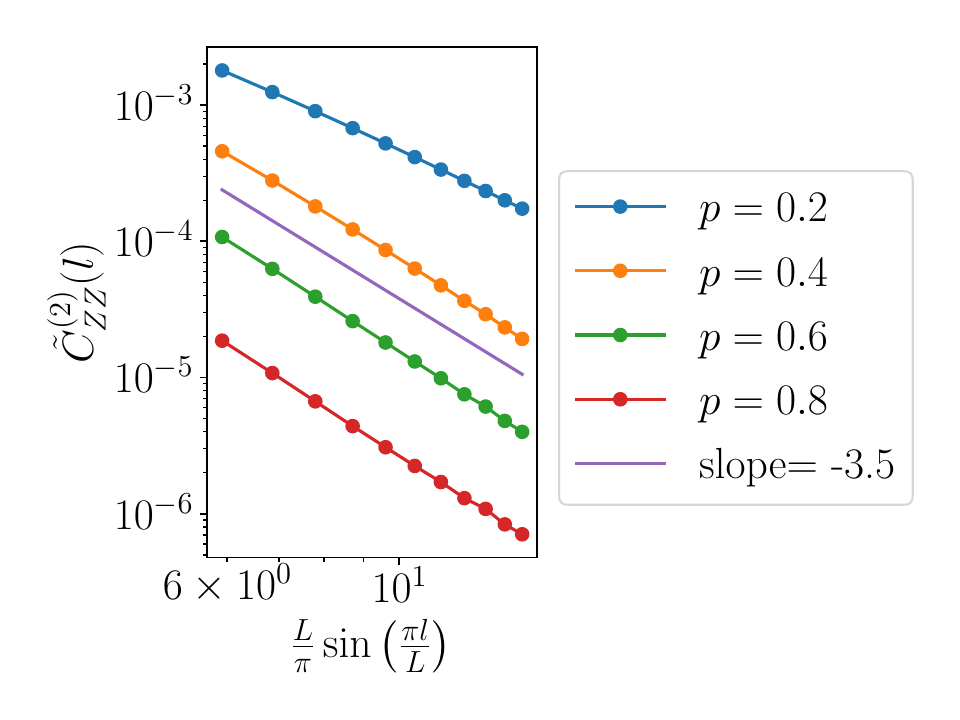}
    \caption{Renyi-2 ZZ correlator $C^{(2)}_{ZZ}(l)$ of the transverse field Ising ground state under X dephasing. For any $0<p<1$ the correlator saturates to a constant at large $l$ (left). After subtracting the constant part, the tail is a power law decay with an exponent $\eta_2 \approx 3.5(2)$ (right).}
    \label{fig:ZZ-X}
\end{figure}

Next we consider the Renyi-2 $ZZ$ correlator
\begin{equation}
    C^{(2)}_{ZZ}(l) = \frac{\tr(\rho^2 Z_0 Z_l)}{\tr(\rho^2)}.
\end{equation}
This again saturates to a constant at long distance, see Fig.~\ref{fig:ZZ-X}. We again find that $C^{(2)}_{ZZ}(l) $ has the same form, which is a constant piece $A$ plus a power-law decaying piece $\tilde{C}^{(2)}_{ZZ}(l)$. We also numerically find that 
\begin{equation}
    A = \left(\frac{\tr(\rho^2 Z_0)}{\tr(\rho^2)}\right)^2, ~~ \mathrm{for~} ZZ \mathrm{~corrlelator}
\end{equation}
which gives
\begin{equation}
    \tilde{C}^{(2)}_{ZZ}(l) = \frac{\tr(\rho^2 Z_0 Z_l)}{\tr(\rho^2)} - \frac{\tr(\rho^2 Z_0)}{\tr(\rho^2)} \frac{\tr(\rho^2 Z_l)}{\tr(\rho^2)}
\end{equation}
The connected correlator decays as
\begin{equation}
    \tilde{C}^{(2)}_{ZZ}(l) = B \left( \frac{L}{\pi}\sin\left(\frac{\pi l}{L}\right)\right)^{-\eta_2}
\end{equation}
where we fit $\eta_2 \approx 3.5(2)$ for $p=0.6$ and $0.8$. The exponent fitted with $p=0.2$ and $p=0.4$ is drifting towards $\eta_2$ at large $l$ as result of the RG flow of the channel towards the $p=1$ fixed point. 

Previously, it was conjectured \cite{ma2023exploringcriticalsystemsmeasurements}  that the defect is factorized into $(|+\rangle\langle +|)^{\otimes 2} + (|-\rangle\langle -|)^{\otimes 2}$, where $|+\rangle$ and $|-\rangle$ are the boundary state of the fixed boundary conditions in the Ising model. It was shown that this gives the correct subleading term in the Renyi entropy $S^{(n)}(\rho_A):=\frac{1}{1-n}\log\mathrm{\tr}(\rho^n_A)$, for both $A$ being the whole system and an interval. Naively, this factorization predicts that $\eta_2 = 4$, which equals the surface critical exponent of the extraordinary boundary phase transition of the Ising model. However, our numerical estimations of $\eta_2$ for both Renyi $XX$ and $ZZ$ correlators are far from $4$.

\subsubsection{Z dephasing}
We consider the Renyi-2 $XX$ and $ZZ$ correlators for the state under $Z$ dephasing. The $XX$ correlator decays as a power law
\begin{equation}
    C^{(2)}_{XX}(l) = B\left( \frac{L}{\pi}\sin\left(\frac{\pi l}{L}\right)\right)^{-\eta_2}
\end{equation}
where $\eta_2$ continuously varies with the noise rate $p$. See Fig.~\ref{fig:ZZ-X}. Also, $\eta_2 =  2\Delta_{\sigma} = 1/4$ for $p=0$ and increases with $p$. The numerics indicate that $\eta_2\rightarrow 1.0$ as $p\rightarrow 1$. 

The Renyi-2 $ZZ$ correlator again saturates to a constant at long-distances. We consider the connected correlator $\tilde{C}^{(2)}_{ZZ}(l)$, where the constant piece has been subtracted, and find a power-law decay,
\begin{equation}
    \tilde{C}^{(2)}_{ZZ}(l) = B\left( \frac{L}{\pi}\sin\left(\frac{\pi l}{L}\right)\right)^{-\eta_2}
\end{equation}
where we fit the slope $\eta_2\approx 2.0(1)$ for small $p$ ($p\leq 0.7$) using the range of $6\leq l \leq 16$ for the $L=64$ spin chain. For larger $p$, we also fit that $\eta_2 \approx 2.8(1)$ for $p=0.8$ and $\eta_2 \approx 3.3(1)$ for $p=0.9$. However, the slope is drifting to larger values as one fits with larger $l$, indicating that the drifting is a finite-size effect. 

To summarize, we have the following table.
\begin{table}[htbp]
    \centering
    \begin{tabular}{|c|c|c|}
    \hline
         &   Constant piece $A$ &  Exponent, $\eta_2$\\ \hline
        $C^{(2)}_{XX}$, $X$ dephasing & Nonzero & 1.25(5)\\ \hline
        $C^{(2)}_{ZZ}$, $X$ dephasing & $(C^{(2)}_{Z})^2$& 3.5(2) 
         \\ \hline
        $C^{(2)}_{XX}$, $Z$ dephasing & 0 & $0.25\leq \eta_2 \leq 1.00(5)$ \\ \hline
        $C^{(2)}_{ZZ}$, $Z$ dephasing ($p\leq 0.7$)& $(C^{(2)}_{Z})^2$ & 2.0(1) \\ \hline
    \end{tabular}
    \caption{The constant part and the exponent of the Renyi correlators, where $C^{(2)}_{Z} :=\tr(\rho^2 Z_0)/\tr(\rho^2)$. The naive expectation that the $X$ dephasing induces a factorized defect suggests that first two exponents $\eta_2 =4$. This expectation is, however, ruled out by the numerical simulation. The field-theoretic explanation of these exponents remains open. } 
    \label{tab:renyicorr}
\end{table}
\begin{figure}
    \centering
    \includegraphics[width=0.45\linewidth]{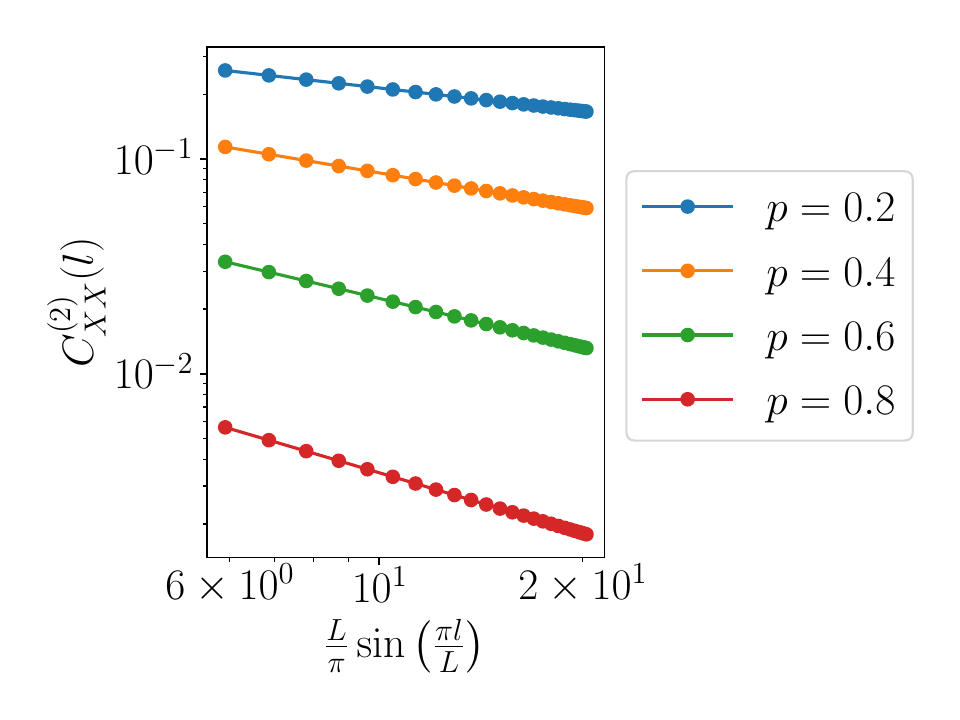}
    \includegraphics[width=0.45\linewidth]{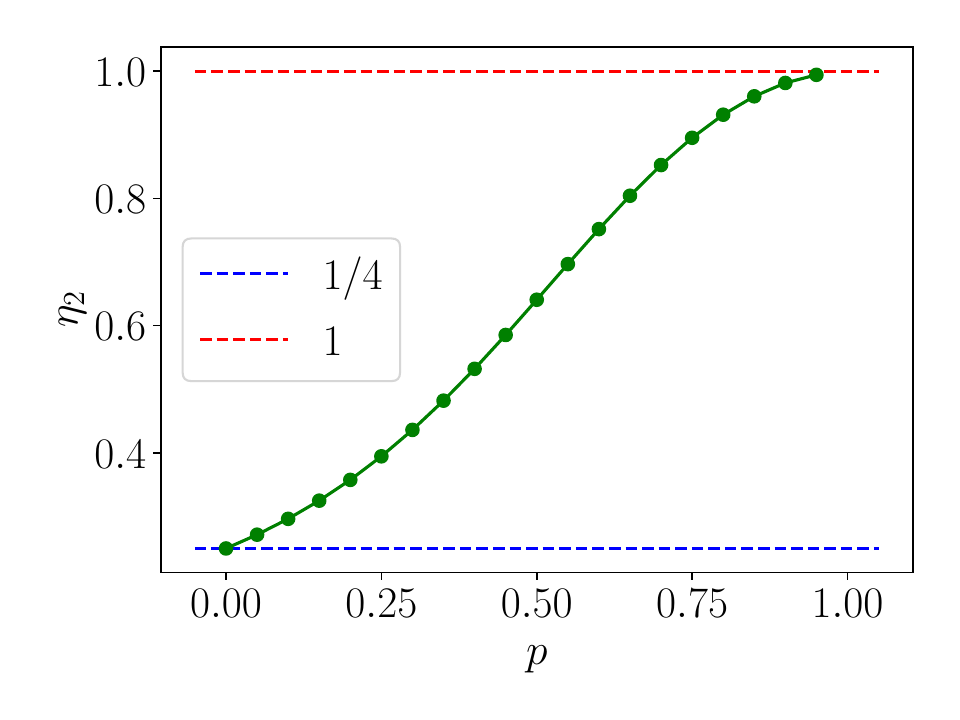}
    \caption{Renyi-2 XX correlator $C^{(2)}_{XX}(l)$ of the transverse field Ising ground state under Z dephasing. The correlator decays as a power law, with exponent $\eta_2$ increasing with $p$ (left). The numerical value is $\eta_2 = 1/4$ at $p=0$ and $\eta_2\rightarrow 1.00(5)$ as $p\rightarrow 1$ (right).}
    \label{fig:XX-Z}
\end{figure}
\begin{figure}
    \centering
    \includegraphics[width=0.45\linewidth]{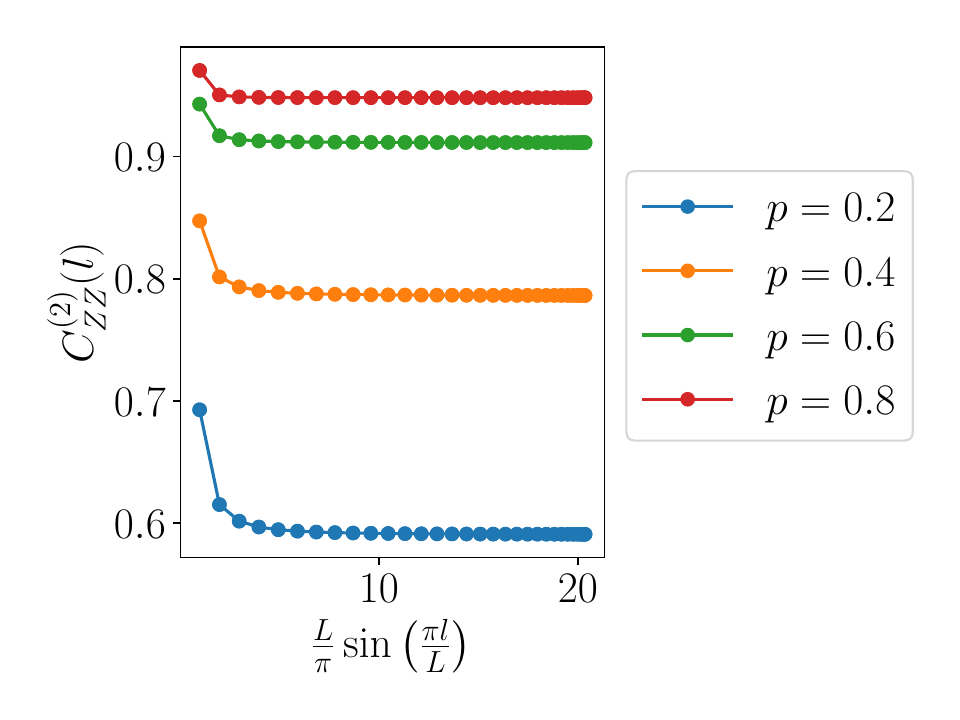}\includegraphics[width=0.45\linewidth]{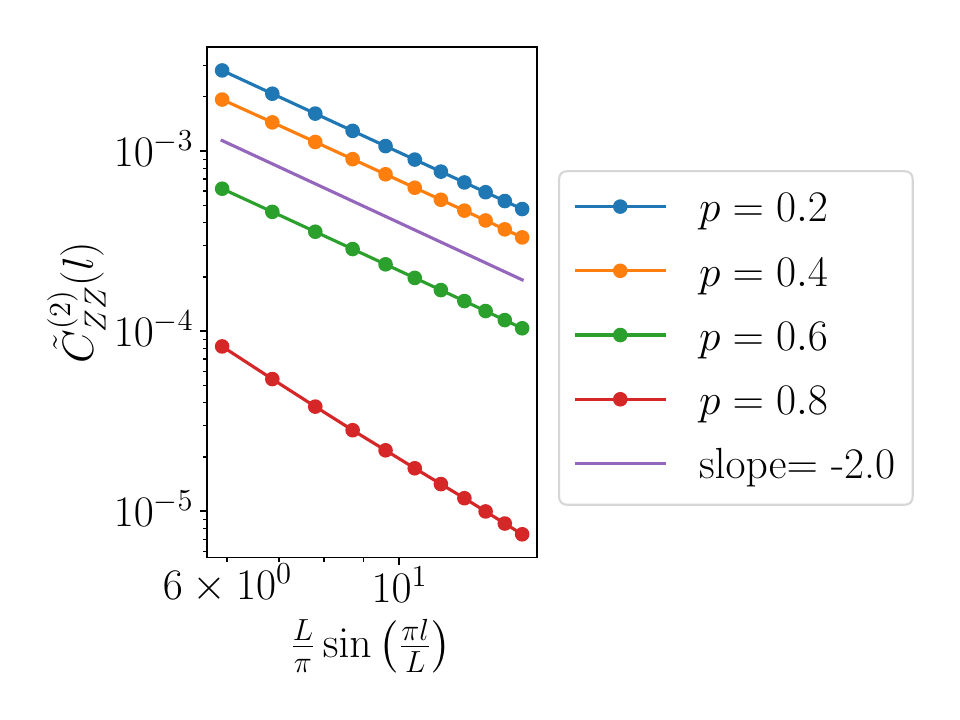}
    \caption{Renyi-2 ZZ correlator $C^{(2)}_{ZZ}(l)$ of the transverse field Ising ground state under Z dephasing. The correlator saturates to a constant at large $l$ (left). After subtracting the constant part, the tail is a power law decay with an exponent $\eta_2 \approx 2.0(1)$ for $p\leq 0.6$ (right). For $p=0.8$, the fit with $6\leq l \leq 16$ gives that $\eta_2\approx 2.8(1)$, which drifts towards larger values for larger $l$. This can be seen as a finite-size effect.} 
    \label{fig:ZZ-Z}
\end{figure}

\section{Optimal fidelity of decoding with different noise strength}

Here, we show the optimal entanglement fidelity from the decoder optimization for $X$ and $Z$ dephasings. We plot the horizontal axis using different exponents $\tilde{p} = pL^{\nu}$, where $\nu = 0,1/2,3/4,1$ to show how $F_e$ changes with $L$ when $\tilde{p}$ is fixed. The theoretical expectation is that for $Z$ dephasing, $F_e$ increases with $L$ as we fix $\tilde{p}$ for whatever $\nu$. For $X$ dephasing, $F_e$ increases with $L$ as we fix $\tilde{p}$ for $\nu>3/4$, and decreases with $L$ for $\nu<3/4$. Due to finite-size corrections, such a decrease can be observed for $\nu=1/2$ at small sizes $L=4,6,8$. 

\begin{figure}
    \centering
    \includegraphics[width=0.95\linewidth]{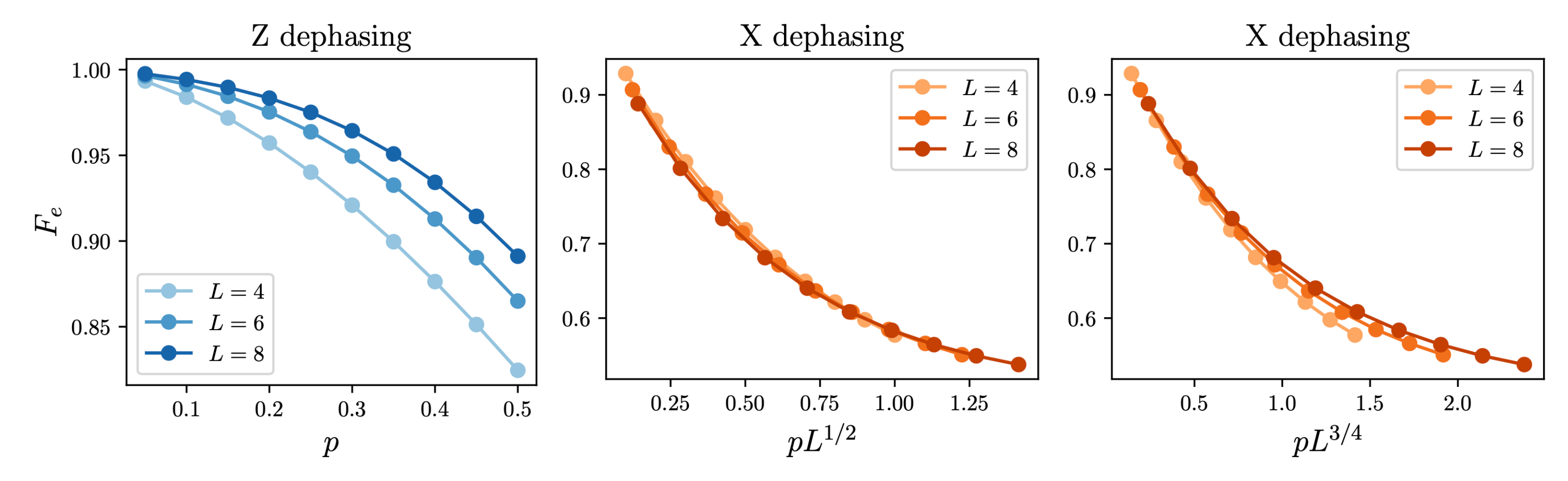}
    \caption{Comparing the $Z$ and $X$ dephasing under different noise rates (from left to right: constant $p$, $p\propto L^{-1/2}$, $p\propto L^{-3/4}$). For $Z$ dephasing and constant $p$, one finds that $F_e$ increases with system size. For $X$ dephasing, the decoder fails to correct the error rate at $p\propto L^{-1/2}$ but can correct a subextensive dephasing error rate at $p\propto L^{-3/4}$.}
    \label{fig:FeX}
\end{figure}

\section{Numerical and experimental methods}
In this section, we explain the numerical and experimental details throughout the paper. 
\subsubsection{Optimal decoder through repeated SVD}
\begin{figure}
    \centering
    \includegraphics[width=0.5\linewidth]{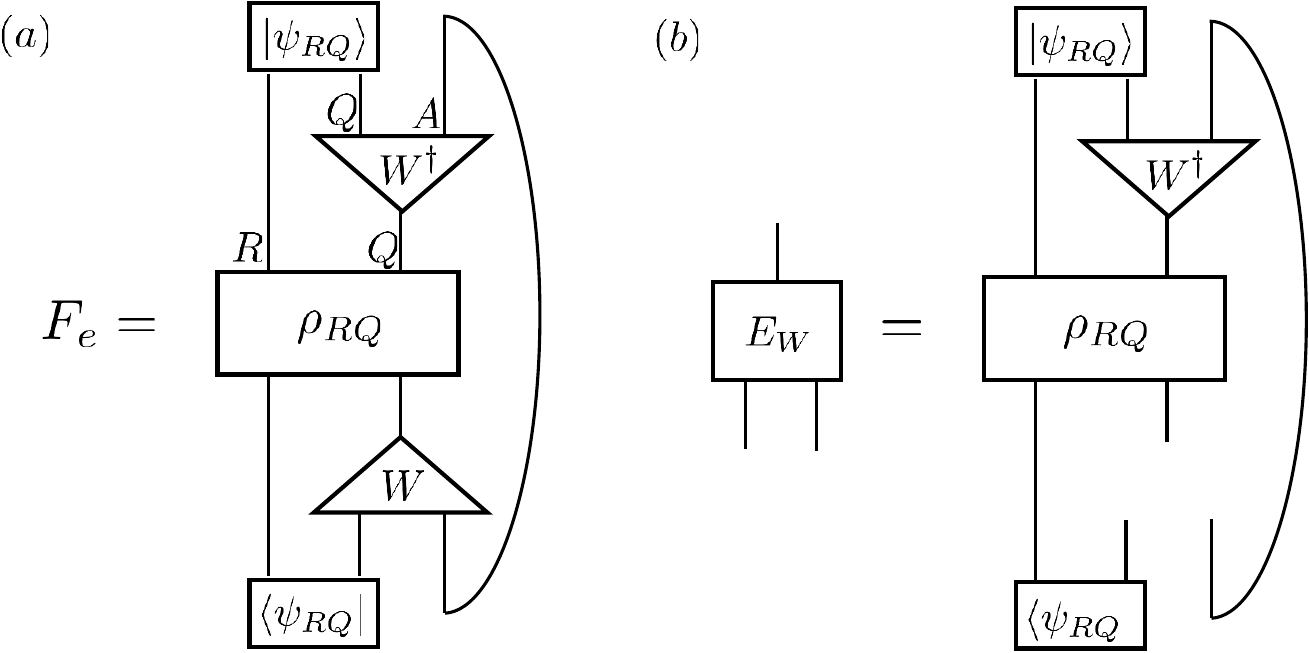}
    \includegraphics[width=0.5\linewidth]{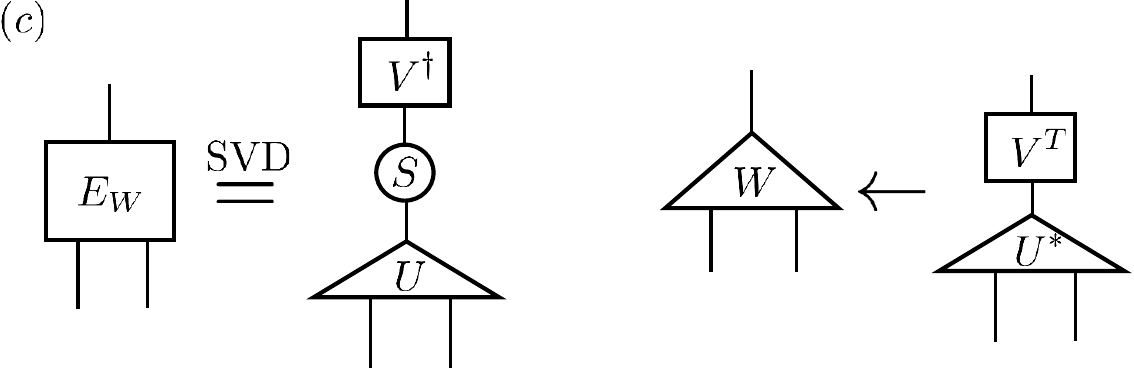}
    \caption{Optimization of the entanglement fidelity $F_e$ through SVD over isometry $W$. (a) Definition of $F_e$. (b) Definition of $E_W$, which gives $F_e = \max_{W}\tr(E^{T}_W W)$. (c) Update rule of $W$ in an iteration.}
    \label{fig:detail_svd}
\end{figure}

Consider the state $|\psi_{RQ}\rangle$ under decoherence $\rho_{RQ} = \mathcal{N}_Q(|\psi_{RQ}\rangle\langle\psi_{RQ}|)$. The entanglement fidelity can be computed as
\begin{equation}
    F_e = \max_{W} \langle \psi_{RQ}|\tr_A(W\rho_{RQ}W^{\dagger})|\psi_{RQ}\rangle
\end{equation}
where $W: Q\rightarrow QA$ is an isometry, $W^{\dagger} W = I$. We optimize $W$ by first rewriting 
\begin{equation}
    F_e = \max_W \tr E^{T}_{W} W, 
\end{equation}
where 
\begin{equation}
    E^{T}_W = \tr_R (\rho_{RQ} W^{\dagger} |\psi_{RQ}\rangle\langle\psi_{RQ}|)
\end{equation}
We can view the tensor $E_W$ as a matrix from $Q$ to $QA$, and perform an SVD,
\begin{equation}
    E_W = U S V^{\dagger},
\end{equation}
where $U^{\dagger}U = I$ and $V^{\dagger}V = I$. The updated $W$ is given by
\begin{equation}
    W \leftarrow U^{*} V^{T}
\end{equation}
and the updated maximum is given by
\begin{equation}
    F_e \leftarrow \tr S.
\end{equation}
We will initialize $W$ with a random isometry and then perform the optimization step above until $F_e$ converges. We find that $F_e$ usually converges to the 5th digit even with 5 iterations. For clarity we also plot $E_W$ and the update rule as tensor network diagrams in Fig.~\ref{fig:detail_svd}. The dimension of the ancilla Hilbert space $A$ can be tuned and we find that $d_A = d_Q$ is sufficient to achieve optimal $F_e$.

The time cost of one iteration consists of the computation of $E_W$, which scales as $O(d^2_R d^2_Q d_A)$, and the SVD step, which scales as $O(d^3_Q d_A)$. We test the case of $L=8$ system qubits and 1 reference qubit on a laptop with Intel i7-10875H (2.3GHZ) processor and find that an iteration costs roughly 8 seconds.
\subsubsection{Gradient-based variational optimization}
To optimize the target function as defined in Eq.\ref{eq:fe}, we treat each two-qubit gate in the ladder circuit as a general SU(4) gate~\cite{zulehner2019compiling} and perform the optimization using PyTorch's automatic differentiation framework, ADAM~\cite{kingma2014adam}. The entanglement fidelity $F_e$ is evaluated by reversing the encoding circuit and measuring the probability in the all-0 output string. 

In the numerical study for the deep variational decoder, we are optimizing a global cost function, the barren plateau issue can be particularly severe, especially when the circuit is deep. To mitigate this issue, we utilize a warm-start strategy, starting the optimization at a shallow depth and iteratively increasing the circuit depth, using the previously optimal parameters as initialization~\cite{truger2024warm,zhang2024scalable}. After reaching the desired depth (presumably in an `overparametrized' regime), we reverse the process to perform a 'scan' over the different number of layers. In practice we find this leads to much better results than a randomized initialization.

\begin{algorithm}[H]
	\caption{A warm-start method for mitigating barren-plateau} 
	\begin{algorithmic}[1]
        \State Set $\tau = 1$
		\While {$\tau<=\tau_{\rm targ}$}

                \State Minimize the entanglement fidelity, $F_e$
                \State Increase $\tau$ by 1; keep the previously optimized parameters initialize the newly added parametrized gates to $I$
		\EndWhile
		\While {$\tau>0$}
                \State Minimize $F_e$
                \State Reduce $\tau$ by 1; keep the previously optimized parameters in the variational circuit
		\EndWhile
	\end{algorithmic} \label{algo:warm}
\end{algorithm}

\subsubsection{Quantinuum H1 Processor}
The Quantinuum System Model H1 is a trapped-ion quantum computer featuring 20 physical qubits arranged in a linear chain. Utilizing a quantum charge-coupled device (QCCD) architecture, it enables all-to-all qubit connectivity by transporting ions between five parallel interaction zones, allowing for flexible and efficient gate operations. The system supports a native gate set that includes single-qubit rotations, two-qubit ZZ gates, and arbitrary-angle ZZ gates. Notably, it offers mid-circuit measurement and qubit reuse capabilities, facilitating advanced quantum algorithms and error correction protocols~\cite{pino2021demonstration,decross2024computational}. 

The H1 achieves high operational fidelity, with single-qubit gate infidelities as low as $1\times10^{-5}$ and two-qubit gate infidelities around $1\times10^{-3}$. State preparation and measurement (SPAM) errors are typically around $3\times10^{-3}$, and memory errors per qubit per depth-1 circuit are approximately $2\times10^{-4}$. 

Based on the above specs, the dominating error factor comes from entangling gates. For decoding of the $L = 8$ CFT code, we use a total of 93 entangling gates: 8 SU(4) gates for each encoding and decoding, additional 15 gates for the decoder; each SU(4) gates can be decomposed into 3 native ZZ gates; the overall estimated infidelity is at $9\%$, matching the observed deviation in Fig.~\ref{fig:fid_exp}.

\subsubsection{Implementation of quantum randomness}
For implementing both the depolarizing channel and the randomized measurement, randomness plays a crucial role. However, QSAM does not have a built-in function for generating random numbers in real time. In theory, one could pre-generate all possible randomized circuits and submit them in advance. However, in practice, this approach introduces significant overhead in input size and increases quantum resource usage, as each quantum circuit submission incurs a fixed cost, regardless of the number of circuit executions.

To address this issue, we utilize extra ancilla qubits along with classical feedback as a natural quantum source of randomness. To implement a Pauli depahsing channel on a physical qubit, we apply a single-qubit rotation \( R_X(\theta) \) to an ancilla qubit, where \( \theta = 2\arcsin(\sqrt{p/2}) \). Upon measuring the ancilla qubit, if the outcome is \( 1 \), a corresponding Pauli gate is applied to the system qubit. Since the QCCD architecture supports mid-circuit measurement and qubit reuse~\cite{pino2021demonstration}, a single ancilla qubit is sufficient for the entire physical system.

Realizing randomized Pauli measurements is slightly more complicated, as we need to select one of the three measurement bases (X, Y, or Z) with equal probability while relying only on qubits instead of qutrits. This can be achieved using two ancilla qubits initialized as 00. Specifically, we define measurement outcomes \( 00 \) and \( 01 \) as corresponding to the Z basis, \( 10 \) to the X basis, and \( 11 \) to the Y basis. The key trick is to apply rotations with different angles: we apply \( R_X(2\arcsin(\sqrt{2/3})) \) to the first ancilla qubit and a Hadamard gate to the second. It is straightforward to verify that the resulting probabilities satisfy \( p_{00} = p_{01} = 1/6 \) and \( p_{10} = p_{11} = 1/3 \), ensuring the desired probabilities of the basis selection.

This method effectively integrates randomness generation into the quantum circuit itself, reducing classical pre-processing overhead while maintaining efficient quantum resource utilization.

\section{Classical shadow estimation of Renyi negativity}
Another previously proposed probe of the mixed-state phases is the Rényi negativity. One such quantity is the Renyi negativity. Given a $L$-qubit, bi-partite quantum system, $AB$, consisting of $|A|=l$ and $|B|=L-l$ spins respectively, the $n$-th order Renyi negativity is defined as the following: $$\mathcal{N}_A^{(n)} = \frac{1}{1-n}{\rm log}[{\rm Tr}(\rho^{T_A})^n/{\rm Tr}(\rho^n)]$$. Without dephasing, the Renyi negativity satisfies the logarithmic scaling,
\begin{equation}
    N^{(3)}_A = \alpha \log \left( \frac{L}{\pi}\sin  \frac{\pi l}{L} \right)+ O(1),
\end{equation}

For the Ising model considered in this work, the Renyi negativity shows distinct features for $X, Y$, and $Z$ dephasings. The Renyi negativity becomes area law for $X$ dephasing, indicating a relevant perturbation. For $Z$ dephasing, the Renyi negativity is still logarithmic, but the coefficient $\alpha$ continuously decreases with $p$, indicating a marginal deformation. For $Y$ dephasing, the Renyi negativity stays as the same logarithmic form with $\alpha=2c/9 = 1/9$. One drawback of this quantity is that it is only nontrivial starting from Renyi index $n\geq 3$. For $n=2$, even though the mixed-state phases are expected to be the same, the Renyi negativity does directly probe the phases.

\begin{figure}
    \centering
    \includegraphics[width=0.35\linewidth]{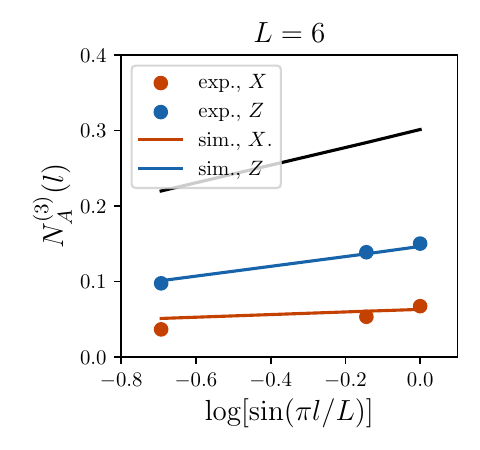}
    \caption{The same experimental data set we use to compute Renyi correlator in the main text can be used to generate the $N_A^{(3)}$ here.}:
    \label{fig:neg}
\end{figure}

The quantity can also be experimentally accessed with the random measurement method~\cite{elben2020mixed}, which we briefly review here. For each copy of the state, we first perform randomized measurement by applying random local unitaries $u_{1}\otimes...u_{L}$, each sampled i.i.d. from a 3-design, and record the classical measurement outcomes $\bm{k} = \{k_1,...k_N\}$. We repeat the the experiment $M$ times and denote each set of measurement outcomes $\bm{k}^{(r)}$, from which we could construct an unbiased estimator:
\begin{align}
    \hat{\rho}^{(r)}_{AB} = \bigotimes_{i\in AB} \hat{\rho}^{(r)}_{i} = \bigotimes_{i\in AB} (3 u_i^{\dag}\ketbra{\bm{k}^{(r)}_i}{\bm{k}^{(r)}_i}u_i - \mathbbm{I})
\end{align}
whose expectation gives raise to $\rho_{AB}$~\cite{huang2020predicting}.  Next, the partially transposed (PT) moments, $p_n = {\rm Tr}[(\rho_{AB}^{T_A})^n]$ can be calculated:
\begin{align}
    p_n & =\frac{1}{n!}{M\choose n}^{-1}\sum_{r_1\neq r_2\neq r_3...\neq r_n}{\Tr{\buildrel \rightarrow \over{\Pi}_A \buildrel \leftarrow \over{\Pi}_B\hat{\rho}^{(r_1)}_{AB}\otimes...\otimes\hat{\rho}^{(r_n)}_{AB}}}\\
    & = \frac{1}{n!}{M\choose n}^{-1}\sum_{r_1\neq r_2\neq r_3...\neq r_n}{\Pi_{i \in A} \Tr{ \hat{\rho}^{(r_1),T}_{i}\otimes...\otimes\hat{\rho}^{(r_n),T}_{i}}}{\Pi_{i \in B} \Tr{ \hat{\rho}^{(r_1)}_{i}\otimes...\otimes\hat{\rho}^{(r_n)}_{i}}} \label{eq:PT}
\end{align}
Notice that the condition $r_1\neq r_2\neq r_3...\neq r_n$ over the sum is necessary for guaranteeing the independence of the samples and forming an unbiased estimation. Naively, the classical computation time required is $O(M^n)$, which is still demanding, considering $M \sim 10^5$ in~\cite{elben2020mixed}. Luckily, one could further simplify Eq.~\ref{eq:PT} due to the symmetries in the equation. For example, for $n = 3$, 
\begin{align}
    p_3 =& \frac{1}{3!}{M\choose 3}^{-1} \tr{\sum_r[\hat{\rho}_{AB}^{(r),T_A}]^3 - 3\sum_r[\hat{\rho}_{AB}^{(r),T_A}]\sum_r[(\hat{\rho}_{AB}^{(r),T_A})^2])} 
    \\&+ 2\tr{\sum_r[(\hat{\rho}_{AB}^{(r),T_A})^3])}
\end{align}
The experimental result for the $L=6$, $p=0.3$ result is demonstrated in Fig.~\ref{fig:neg}. The difference between $X$ and $Z$ channels are clearly captured, in accordance with the theoretical expectation.

\end{document}